# Ultra-short lifetime isomer studies from photonuclear reactions using laser-driven ultra-intense γ-ray

Di Wu[1,2†], Haoyang Lan[1,2†], Jiaxing Liu[1,2†], Huangang Lu[1,2†], Jianyao Zhang[1,2], Jianfeng Lv[1,2], Xuezhi Wu[1,2], Hui Zhang[1,2], Yadong Xia[1,2], Qiangyou He[1,2], Jie Cai[1,2], Qianyi Ma[1,2], Yuhui Xia[1,2], Zhenan Wang[1,2], Meizhi Wang[1,2], Zhiyan Yang[1,2], Xinlu Xu[1,2], Yixing Geng[1,2], Chen Lin[1,2], Wenjun Ma[1,2], Yanying Zhao[1,2], Haoran Wang[3], Fulong Liu[3], Chuangye He[3], Jinqing Yu[4], Bing Guo[3], Guoqiang Zhang[5], Furong Xu[1], Naiyan Wang[3], Yugang Ma[6], Gérard Mourou[2,7], Xueqing Yan[1,2*]

**Affiliations:**

[1]State Key Laboratory of Nuclear Physics and Technology, School of Physics, CAPT, University of Wherever, Peking University, Beijing 100871, China.

[2]Beijing Laser Acceleration Innovation Center, Beijing 101407, China.

[3]Department of Nuclear Physics, China Institute of Atomic Energy, Beijing 102413, China.

[4]School of Physics and Electronics, Hunan University, Changsha 410012, China.

[5]Shanghai Advanced Research Institute, Chinese Academy of Sciences, Shanghai 201210, China.

[6]Key Laboratory of Nuclear Physics and Ion-beam Application (MOE), Institute of Modern Physics, Fudan University, Shanghai 200433, China.

[7]IZEST, Ecole Polytechnique, 91128 Palaiseau, France

*Corresponding author. Email: x.yan@pku.edu.cn

†These authors contributed equally to this work.

**Abstract:** Isomers, ubiquitous populations of relatively long-lived nuclear excited states, play a crucial role in nuclear physics. However, isomers with half-life times of several seconds or less barely had experimental cross section data due to the lack of a suitable measuring method. We report a method of online γ spectroscopy for ultra-short-lived isomers from photonuclear reactions using laser-driven ultra-intense γ-rays. The fastest time resolution can reach sub-ps level with γ-ray intensities $>10^{19}$/s ($\geq 8$ MeV). The $^{115}$In(γ, n)$^{114m2}$In reaction ($T_{1/2} = 43.1$ ms) was first measured in the high-energy region which shed light on the nuclear structure studies of In element. Simulations showed it would be an efficient way to study $^{229m}$Th ($T_{1/2} = 7$ μs), which is believed to be the next generation of nuclear clock. This work offered a unique way of gaining insight into ultra-short lifetimes and promised an effective way to fill the gap in relevant experimental data.

**One-Sentence Summary:** Laser-driven ultra-intense γ-rays can be used to measure cross section data of short-lived isomers from photonuclear reactions. The time resolution can reach sub-ps level with γ-ray intensities $>10^{19}$/s ($\geq 8$ MeV), which can produce a decent yield of reaction product. The $^{115}$In(γ, n)$^{114m2}$In reaction with a half lifetime of 43.1 ms was first measured in the high-energy region which shed light on the nuclear structure studies of In element. It can be also used as an efficient pump for $^{229m}$Th nuclear clock and provide a unique way for short-lived isomer studies which extremely lack experimental data.

Isomers, as a bridge between nuclear and atomic physics (*1*), are nuclei in relatively long-lived 'meta-stable' excited states (*2*). Nuclear isomerism has made great progress in the past decades (*2,3*) since it was first put forward by Soddy in 1917 (*4*). More than 1318 isomers are discovered with half-lives ranging from nanoseconds to years and energies ranging from 8.3 eV ($^{229m}$Th, $T_{1/2}$ = 7 μs) (*5,6*) to 13.8 MeV ($^{208m}$Pb, $T_{1/2}$ = 60 ns) (*7*), suggesting that isomers are important populations of nuclear excited states. A series of ultra-short lifetime isomers are proved to be of great interest in nuclear structures (*8-11*) and nuclear astrophysics (*12-14*) and have great application prospects in nuclear laser (*15*), nuclear quantum optics (*16*), and nuclear clock (*6,17*). For even-even nuclei, $I^\pi = 0^+$ isomers are all short-lived with a $T_{1/2} < 3$ μs (*18*). Recently, a fission isomer of $^{235}$U with a $T_{1/2}$ = 3.6 ms was identified (*19*). For nucleosynthesis (*14*), isomer states directly change the reaction threshold energy, they will not only provide extra sensitivities but also influence the nucleosynthesis itself (*20*). Short-lived isomers have a significant impact on the *r*-process (*13*) and *p*-processes (*12,21*), which mostly happened in the Type II Supernova explosion. However, isomers from photonuclear reactions with half-life times of several seconds or less barely had experimental cross section data. The short lifetimes make the offline activation method very difficult and the online photoneutron measurement cannot distinguish neutron reaction channels. To investigate those ultra-short lifetime isomers, ultra-bright ultra-fast γ-ray sources are required.

The laser-driven bremsstrahlung γ-ray source, generated by laser wakefield accelerated (LWFA) monoenergetic electrons (*22-25*), can provide a stable ultra-bright, ultra-fast, and energy-adjustable photon beam (*26-29*), which is suitable for photonuclear reaction studies. The intensities of laser-driven γ-ray are several orders of magnitude stronger than other γ-ray sources, it is more favorable for the reaction measurements with very short half-lives. In addition, the in-beam background radiation for laser-driven nuclear reactions is highly related to the laser pulse, the duration only lasts at a nanosecond level, which can provide a very low background environment for online activation measurements and significantly improve the signal-background ratio.

In our previous works, a laser-driven ultra-bright ultra-fast γ-ray source was established using LWFA electrons with instantaneous intensities above $10^{19}$/s ($10^{7\sim8}$ per shot), photonuclear reactions of $^{197}$Au(γ, xn; x = 1~9) (*30*), $^{92}$Mo(γ, n) $^{91m,g}$Mo (*31*), and some other isotopes were measured by offline activation method. The results showed good agreement with previous works using traditional γ-ray sources and proved this laser-driven γ-ray source to be accurate.

A unique method of online γ spectroscopy for ultra-short lifetime isomers from photonuclear reactions using laser-driven ultra-intense γ-ray was provided in this paper. The $^{115}$In(γ, n) $^{114m2}$In reaction, with a half-life time of 43.1 ms, was first measured in the high-energy region. A potential photonuclear reaction pumping of $^{229m}$Th nuclear clock was first proposed by GEANT4 (*32,33*) simulation. This work offered a unique way of gaining insight into photonuclear reaction research with ultra-short lifetimes, which extremely lacks experimental work. The fastest time resolution can reach sub-ps level due to the ultra-short duration time of laser-driven γ-ray, and the intensities of $10^{7\sim8}$ per shot can provide a decent yield for reactions like (γ, xn), (γ, xp), (γ, α), even (γ, fission).

**Laser-driven ultra-intense γ-ray**

The experiment was carried out at the 200 TW laser facility in the Compact Laser Plasma Accelerator Laboratory (CLAPA), Peking University. The schematic layout of the experimental setup is shown in Fig. 1 (a). A supersonic nozzle, providing a supersonic expansion with a maximum air pressure of 45 bar (plasma density about $1*10^{19}$ cm$^{-3}$), was used to generate LWFA monoenergetic electrons. The plasma intensities were diagnosed with a probe light by interference imaging (Fig. S1 (a)). An Al foil was used to reflect the rest of the laser and a Be window was used to seal the vacuum in the electron path. The electron charges of each shot were measured and recorded by a Turbo Integrating Current Transformer (Turbo-ICT) (*34*), while the electron spectra were analyzed by a magnetic spectrometer. Detailed electron parameters are shown in Fig. S1 (b), (c), (d). The bremsstrahlung γ-ray was generated by a 2 mm Ta (99.99%) converter. The γ-ray energy spectra were obtained by GEANT4 simulations (Fig. 1 (b)) and the intensities were given by the Turbo-ICT.

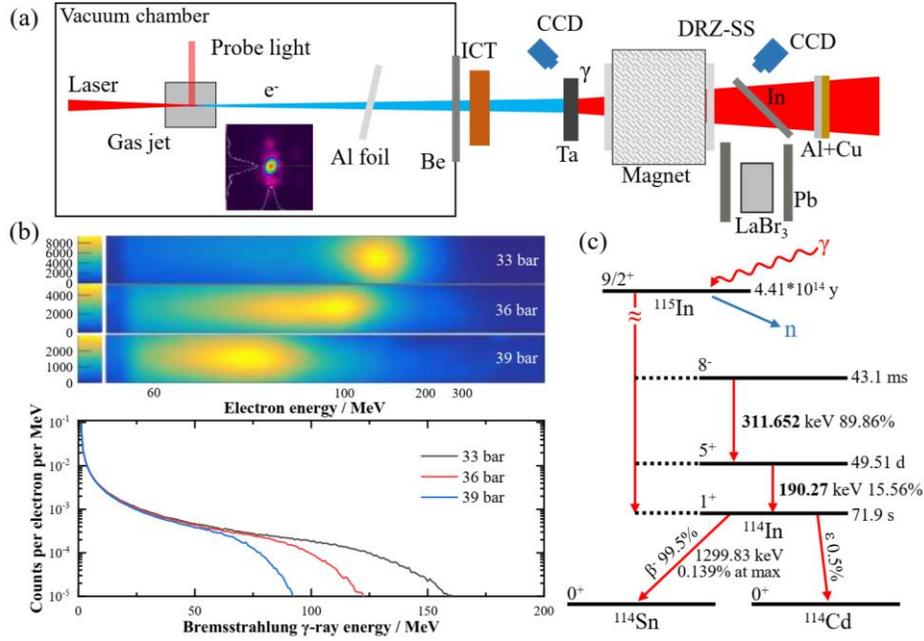

**Fig. 1** (a) Schematic layout of experimental setup. The focal laser spot with a size of 21*23 μm full width at half maximum is also shown in the figure. (b) Typical LWFA electron spectra obtained by the magnetic spectrometer at gas pressure of 33, 36, 39 bar, respectively. The bremsstrahlung γ-ray energy spectra were obtained by GEANT4 simulations using the averaged electron spectra of 100 continuous shots. (c) Relative nuclear decay data of the radioactive nuclei from the $^{115}$In(γ, n) $^{114}$In reaction. The ground state cannot be measured due to its lack of suitable γ-ray decays.

**Short-lived isomer measurements**

A laminated target made of Al and Cu was used as γ-ray flux monitors for each electron energy. The $^{nat}$In (99.99%, $^{115}$In 95.71%) targets had a size of 2.8*2.0 cm and a thickness of 1 mm, it was set at 45° along the long side to the laser axis. The relative nuclear decay data of the radioactive nuclei from the $^{115}$In(γ, n) $^{114}$In reaction were listed in Fig. 1 (c), which were taken from NuDat 3.0 database (*35*).

A 3*3 inch LaBr$_3$ detector was used to measure the $^{114m2}$In decay signals at 90° to the laser axis (the opposite direction of the remaining electrons). The acquisition time of the LaBr$_3$ detector was controlled by gate signals to reduce the influence of background radiation and secondary radiation of the electron beams. A typical online spectrum of 800 shots at gas pressure of 33 bar was shown in Fig. 2, the 312 keV γ-ray decay of $^{114m2}$In was clearly distinguished. The half-life time was analyzed to be 43.7(1.1) ms, which matched well with the previous value, this result further verified the source of the 312 keV γ-ray signals. The 511 keV online background was caused by the positron annihilation produced by the high energy γ-ray, and an online background radiation of around 200 keV from unknown sources was also found in the spectrum. In order to obtain a better net count, the online background (Fig. 3) at each electron energy was measured for 400 shots, the 312 keV signal disappeared while the others remained. After online measurement, the $^{114m1}$In was measured using an HPGe detector with a relative efficiency of 40%, a typical spectrum was shown in Fig. S2.

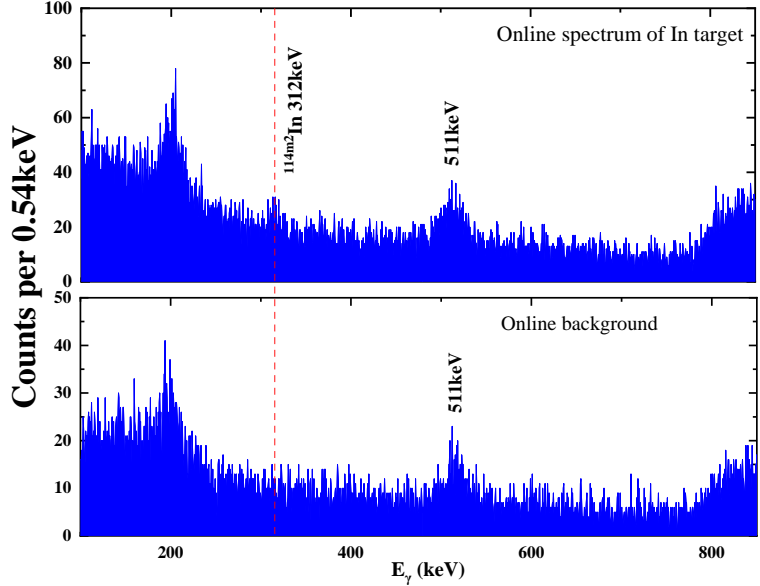

**Fig. 2** Typical online spectrum of In target and background at gas pressure of 33 bar.

The flux-weighted average cross sections (FACS) of $^{115}$In(γ, n) $^{114m1,m2}$In reactions were listed in Tab. 1, and compared with previous works (*36,37*) and TALYS 1.9 calculation (*38*) in Fig. 3. Our results of the $^{115}$In(γ, n)$^{114m1}$In / $^{197}$Au(γ, n)$^{196}$Au yield ratios matched well with previous work (*39*), which were slightly lower than TALYS 1.9. But for $^{115}$In(γ, n)$^{114m2}$In reaction, TALYS 1.9 calculation is 48% less than our experiment results at maximum. This discrepancy between theory calculations and experiments might be caused by the inappropriate parameters of the nuclear reaction model, nuclear level density, or optical potential, more theory calculations are still needed for this reaction.

Table 1: Experimental flux-weighted average cross sections of $^{115}$In(γ, n)$^{114}$In reaction determined by present work and TALYS 1.9 calculations using monoenergetic electron beams at 78 MeV, 103 MeV, and 135 MeV.

| Reaction | Center energy of electrons / MeV | Experimental cross section / mb | TALYS 1.9 / mb |
|---|---|---|---|
| $^{115}$In(γ, n)$^{114m1}$In | 78±10 | 25.3±1.7 | 26.3 |
| | 103±14 | 21.4±1.7 | 23.4 |
| | 135±20 | 19.8±1.8 | 21.3 |
| $^{115}$In(γ, n)$^{114m2}$In | 78±10 | 12.7±1.0 | 8.7 |
| | 103±14 | 11.3±1.0 | 7.8 |
| | 135±20 | 10.6±1.1 | 7.2 |

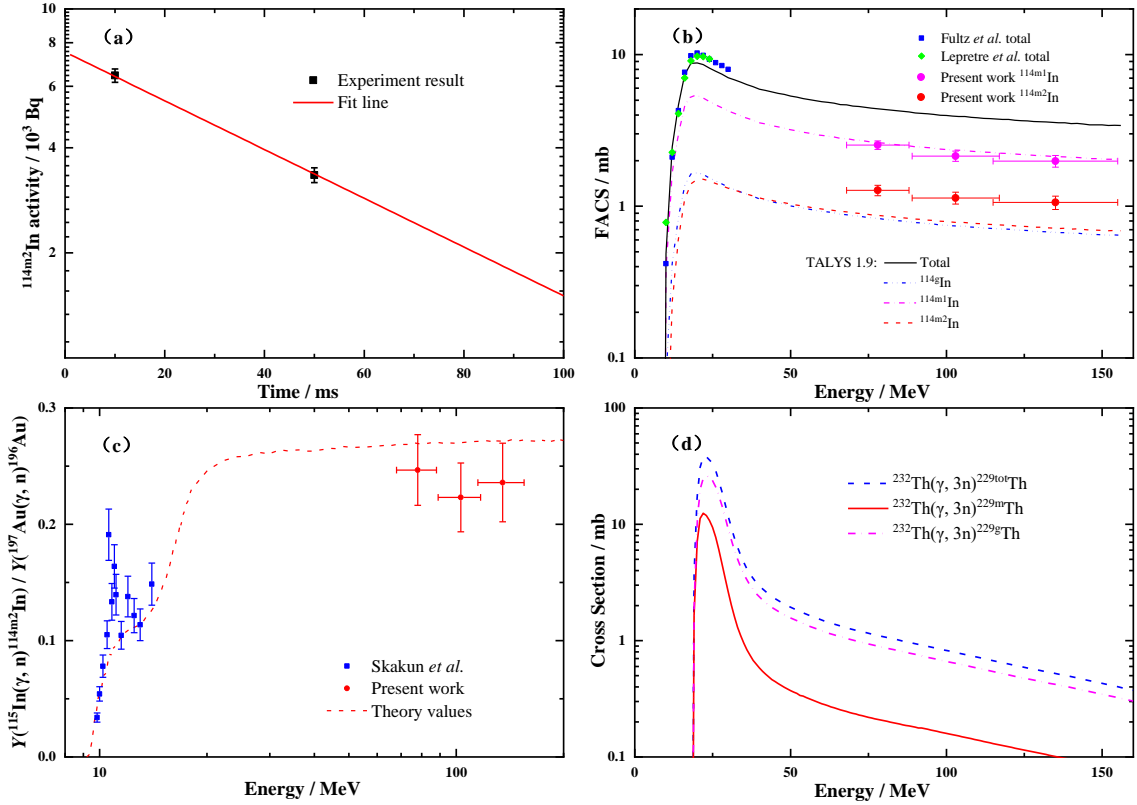

**Fig. 3** (a) Decay analysis of the $^{114m2}$In. (b) Experimental results of $^{115}$In(γ, n)$^{114}$In reaction FACSs, comparing with previous works and TALYS 1.9 calculations. (c) Experimental ratios of $^{115}$In(γ, n)$^{114m2}$In/$^{197}$Au(γ, n)$^{196}$Au yield at per target atom, comparing with previous work and theory values simulated by GEANT4 code using TALYS 1.9 calculations. (d) $^{232}$Th(γ, 3n)$^{229}$Th reaction cross sections from TALYS 1.9.

## $^{229m}$Th nuclear clock pumping and more short-lived isomer studies

This online γ spectroscopy method using laser-driven ultra-intense γ-ray was proved to be suitable for short-lived isomers studies in photonuclear reactions. With proper fast nuclear electronics (*40*) and a more reasonable experimental setup, ultra-short lifetime isomers with $T_{1/2}$ at tens of picoseconds or less could be directly measured. The laser-driven γ-ray intensities are several orders of magnitude stronger than other γ-ray sources such as laser Compton scattering (LCS) (*41,42*) or electron accelerator bremsstrahlung (*43*) (Fig. 4 (a)), which can generate relatively more effective yields for ultra-short-lived isomers and is a great advantage for improving signal-to-noise ratio in online measurement.

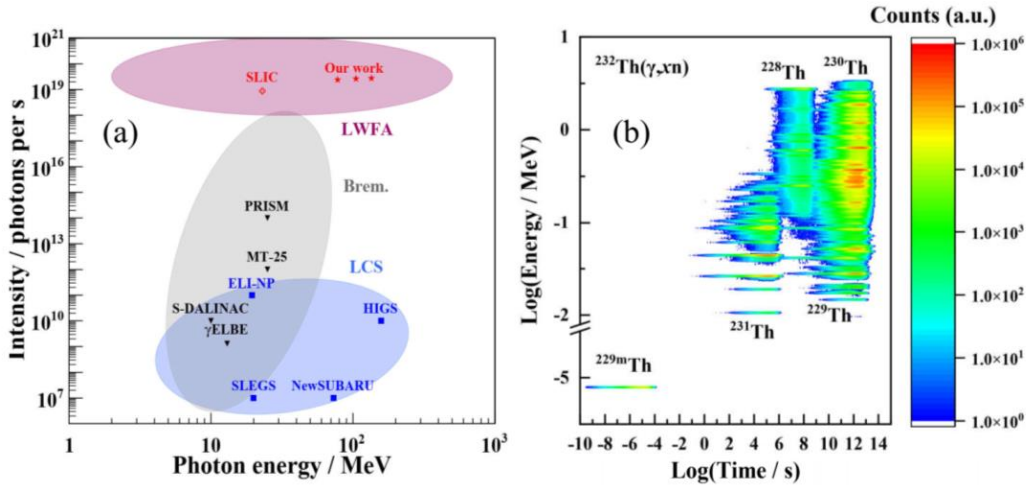

**Fig. 4** (a) A comparison of the peak γ-ray intensities of different facilities. The present laser-driven γ-ray intensities are about 5 orders of magnitude stronger than electron accelerator bremsstrahlung and about 9 orders of magnitude stronger than LCS γ source at high energy region. (b) The correlation between the transition energy and the detection time for the radioactive products from the $^{232}$Th(γ, xn) reactions using GEANT4 simulation. A decent yield of $^{229m}$Th can be generated as a potential pumping for nuclear clock. The $^{229m}$Th decay was assumed to be a purely direct photon transition with a half-life of 7μs in the simulation.

Fig. 4 (b) demonstrated that $^{229m}$Th, a potential nuclear clock (*6,17*), can be produced by the $^{232}$Th(γ, 3n) reaction, therefore, the laser-driven $^{232}$Th(γ, 3n)$^{229m}$Th reaction can be used as a pump for a new generation nuclear clock. This pumping scheme has several advantages. Firstly, the ultra-short nature of the laser-driven γ-ray source can be utilized to match the very short lifetime of $^{229m}$Th. Secondly, the natural $^{232}$Th targets can be used instead of artificially manufactured $^{229}$Th, which is more cost-effective. Thirdly, with the high acceleration gradient of laser-plasma accelerators, such isomer pumping devices can be made compact and more easily accessible. As shown in Tab. S1 and Fig. S2, plentiful short-lived isomers can be produced via laser-driven photonuclear reactions. The relatively long-lived (sub μs) $0^+$ even-even nuclei such as $^{72m}$Ge and $^{98m}$Ge can also be studied using this method (*18*). More experiments on ultra-short lifetime isomers will be performed in the future, promising an effective way to fill the gap in relevant experimental data.

**Acknowledgments:** The authors thank the staff of the 200 TW laser of CLAPA laboratory for the smooth operation of the machine.

**Funding:**

National Natural Science Foundation of China grant 11921006, 12305266, 12125509

Beijing Outstanding Young Scientists Program

National Grand Instrument Project grant 2019YFF01014400

Open Foundation of Key Laboratory of High Power Laser and Physics, Chinese Academy of Sciences grant SGKF202104

**Author contributions:**

Conceptualization: DW, HYL, JXL, HGL

Methodology: DW, HYL, JXL, HGL, JYZ, JFL, XZW, HZ, YDX, QYH, JC, QYM, YHX, ZNW, MZW, ZYY

Recourses: XLX, YXG, CL, WJM, YYZ, HRW, FLL, CYH, JQY, BG, GQZ, FRX, NYW, YGM, XQY

Investigation: DW, HYL, JYZ, MZW

Software: DW, HYL, JYZ, MZW

Visualization: DW, HYL

Funding acquisition: DW, BG, XQY

Project administration: GM, XQY

Supervision: XQY

Writing – original draft: DW

Writing – review & editing: HYL, JXL, XQY

**Competing interests:** Authors declare that they have no competing interests.

**Data and materials availability:** Data will be made available on request.

**SUPPLEMENTARY MATERIALS**

Materials and Methods

Figs. S1 to S3

Table S1